\documentstyle[floats,tighten,preprint,prd,aps,epsf]{revtex}

\begin{document}
\preprint{\parbox[t]{15em}{\raggedleft
FERMILAB-PUB-98/139-T \\ hep-ph/9805215\\[2.0em]}}
\draft

% add words to TeX's hyphenation exception list
\hyphenation{author another created financial paper re-commend-ed}

% declarations for front matter
\title{The Perturbative Pole Mass in QCD}

\author{Andreas S. Kronfeld}

\address{Theoretical Physics Group,
Fermi National Accelerator Laboratory, Batavia, Illinois}

\date{1 May 1998}

\maketitle % typeset front matter (abstract goes after for REVTeX)

\widetext
\begin{abstract}
It is widely believed that the pole mass of a quark is infrared-finite
and gauge-independent to all orders in perturbation theory.
This seems not to have been proved in the literature.
A proof is provided here.
\end{abstract}

\pacs{PACS numbers: 12.15.Ff, 11.15.Bt, 12.38Dx}

\narrowtext
\epsfverbosetrue

\section{INTRODUCTION}
Long ago Tarrach~\cite{Tar81} showed that the pole mass of a quark
is infrared-finite and gauge-independent through two-loop order in
perturbation theory.
It is widely believed that this result holds through any finite
order in perturbation theory.
There does not seem to be a reference in the literature containing
a proof, however, and this paper aims to fill that gap.
To be specific I consider QCD with one massive quark and $n_f-1$
massless quarks, but extra massive quarks do not change the argument.

There are a few reasons why one might suspect infrared divergences
to arise in the perturbative-QCD series for the pole mass.
QCD contains massless self-interacting gluons, so the infrared behavior
is often worse than for QED.
In QCD and QED, infrared divergences {\em do\/} arise in the two-loop
self energy.
Another worry is that the {\em non}perturbative pole mass, if at
all defined, is clearly very sensitive to the infrared.

On the other hand, the kinematics of QCD, with the quark content
under consideration, is like that of QED with a massive muon, some
massless scalar bosons, and some massless electrons~\cite{App76}.
Here (or with $n_f$ large enough to make QCD infrared-free) 
nonperturbative problems should not arise in the infrared.
For the QED model one can even pick boundary conditions so that
gauge-invariant muon states are in the spectrum.%
\footnote{The magnetic monopole considered by Ref.~\cite{Pol91} is a
prototype of such a state.}
Also, the infrared divergence from the two-loop self energy is
cancelled by a term from evaluating the one-loop self energy on
the one-loop mass shell and iterating~\cite{Tar81,Gra90}; one could
hope that this mechanism occurs at any order.
Moreover, in asymptotically free QCD a remnant of the anticipated
infrared sensitivity appears through renormalons~\cite{Big94,Ben94};
this implies that perturbation theory can ``know about'' the infrared
behavior of nonperturbative QCD without having infrared divergences
at fixed order.

The above discussion is nothing but a duel of fears and hopes, and
it should be replaced by definitive results.
Below it is shown that the infrared divergences cancel in the pole 
mass, as at two loops, to any order.
It is then a simpler matter to confirm that the pole mass is
independent of the gauge chosen for the calculation.
The arguments are straightforward and can be found in textbooks,
although, to the best of my knowledge, the specific application is
not.

This paper is organized as follows:
Some notation is in Sec.~\ref{Notation}.
Section~\ref{IRF} proves that the pole mass is infrared-finite
at every order in perturbative QCD.
Section~\ref{GI}  proves that the pole mass does not depend on the 
gauge-fixing function through every finite order.
Some concluding remarks are in Sec.~\ref{Conclusions}.

\section{Notation}
\label{Notation}
This paper uses the metric $g_{\mu\nu}=\mathop{\rm diag}(-1,1,1,1)$
in Minkowski space-time.
In particular, the momentum~$p$ of a real particle with mass~$M$
satisfies $p^2=-M^2$.

The pole mass is derived by identifying the pole in the massive quark's
full propagator~$S(p)$.
One has
\begin{equation}
    S^{-1}(p)=i\kern+0.1em /\kern-0.55em p + m_0 - \Sigma(p),
    \label{S(p)}
\end{equation}
where $m_0$ is the bare mass and the self energy~$\Sigma$ is given by
the sum of one-particle irreducible Feynman diagrams.
One can write
\begin{equation}
    \Sigma(p,m_0)=i\kern+0.1em /\kern-0.55em p A(p^2,m_0) + m_0B(p^2,m_0),
    \label{Sigma}
\end{equation}
exhibiting the parametric dependence of~$\Sigma$ on the bare
mass~$m_0$.
In considering the dependence on~$p^2$ it is convenient to use a 
slight abuse of notation,~$\Sigma(p^2,m_0)$.

The propagator $S(p)$ has a pole at $p^2=-M^2$, where
\begin{equation}
    M=m_0Z_m
\end{equation}
and
\begin{equation}
    Z_m=\lim_{p^2\to-M^2}\frac{1-B(p^2,m_0)}{1-A(p^2,m_0)},
    \label{Z}
\end{equation}
provided the limit is not infrared divergent.
In perturbation theory one applies Eq.~(\ref{Z}) by expanding the
right-hand side through~$L$th order and setting~$p^2$ to~$-M^2$
iteratively.
A gauge-invariant ultraviolet regulator, such as dimensional
regularization or a lattice, is assumed but not made explicit.
Because gauge theories are renormalizable, ultraviolet divergences
of the coefficients~$Z_m^{[l]}$ are compensated by the bare mass
and gauge coupling.%
\footnote{For example, in dimensional regularization one could
introduce a renormalized mass~$\bar{m}(\mu)=m_0Z_{\bar{m}}$ by
minimal subtraction.  Then one could focus on the infrared behavior
of~$Z_m/Z_{\bar{m}}$.}

Perturbative series are written, for example, as
\begin{equation}
    Z_m=1+\sum_{l=1}^\infty g_0^{2l}Z_m^{[l]},
    \label{Z expansion}
\end{equation}
where $g_0^2$ is the bare gauge coupling.
Below it is convenient to use a short-hand $[\bullet]^{[l]}$ for
the~$l$th term in the perturbative series of expressions abbreviated
here with~$\bullet$.

The momentum~$p$ is reserved for the external momentum of the quark.
Loop momenta are denoted generically by $k$.

\section{Infrared Finiteness}
\label{IRF}
To prove infrared-finiteness I follow the methods of Chapter~13
of Ref.~\cite{Ste93}.
First, I recall how, in perturbation theory, one finds singularities 
in Green functions.
This analysis establishes that the propagator and the self energy have 
a branch point at $p^2=-M^2$.
Since the pole mass requires the self-energy functions to be evaluated
here, one must check whether they diverge at the branch point or not.
It turns out that the on-shell self energy does suffer from infrared 
divergences, but I show that they drop out of the pole mass.

\subsection{Location of Singularities}
Consider an arbitrary Feynman diagram (of any Green function), with 
quark propagators rationalized and all denominators combined with 
Feynman parameters~$\alpha_i$.
If the diagram has~$n$ lines the resulting denominator is
\begin{equation}
    D=\left[\sum_{i=1}^n \alpha_i(q_i^2+m_i^2)\right]^n,
\end{equation}
where $q_i=q_i(p,k)$ are the momenta of the internal lines.
The Green function is an analytic function of Lorentz invariants of 
the external momenta, up to branch points.
Branch points can arise only when~$D$ vanishes, but that is not 
enough.
In addition, the contour of integration (over Feynman parameters~$\alpha$
and loop momenta~$k$) must be pinched~\cite{Ste93}.
This happens if and only if on each internal line
\begin{equation}
    q_i^2+m_i^2=0~\text{\em or}~\alpha_i=0
    \label{on shell}
\end{equation}
and, furthermore, following any closed path~$\ell$ in the diagram
\begin{equation}
    \sum_{i\in\ell}\alpha_i q_i = 0,
    \label{closed path}
\end{equation}
with the sign of $q_i$ taken in the sense of the path~$\ell$.
Equations~(\ref{on shell}) and~(\ref{closed path}) are the so-called 
Landau equations.

Solutions of the Landau equations have a physical 
interpretation~\cite{Col65}.
Up to an overall factor the Feynman parameter~$\alpha_i$ is the
ratio of the time elapsed, from one end of the line to the other, to
the energy propagating on the line.
Thus,~$\alpha_i q_i$ is the space-time separation between the two
ends, and Eq.~(\ref{closed path}) says a loop in the diagram corresponds
to a loop in space-time.
Furthermore, Eq.~(\ref{on shell}) says that internal lines either are
on shell ($q_i^2=-m_i^2$) or do not propagate ($\alpha_i=0$).
For each diagram one obtains the {\em reduced diagram\/} by shrinking
off-shell lines to a point.
Then branch points arise if and only if the reduced diagram represents 
a genuine physical process of on-shell states.

The physical picture given above is useful, because it is often easier
to find solutions to Eqs.~(\ref{on shell}) and~(\ref{closed path}) with
physical reasoning instead of with algebra.
For example, a two-point function has branch points only at normal
thresholds, that is, when~$p^2$ is just right to produce a collection
of on-shell particles.
For the massive quark propagator these branch points are at
\begin{equation}
p^2=-[(1+2r)M]^2,
\end{equation}
corresponding to creation of the massive quark plus~$r$ massive pairs.
These branch points are accumulations of infinitely many solutions
to the Landau equations, because once a solution is found, others
are given by adding zero-momentum massless lines.
Physically, this is because it costs nothing to create an extra soft
gluon or extra soft pair.%
\footnote{If one of the other quarks had a mass~$M_l$, the branch
points would be at $p^2=-[(1+2r)M+2sM_l]^2$, so infinitely many
collapse to the same point as~$M_l\to0$.}
If the solutions accumulate too quickly, an infrared divergence
will develop.

On the other hand, note that there are no collinear divergences.
As soon as the massive quark radiates non-zero momentum, it is off 
shell, and the (un)physical picture disallows a singularity.

\subsection{Infrared Divergences}
To examine the infrared properties, one performs a power-counting
analysis.
One scales some or all loop momenta by a factor $\lambda$; if the
Feynman integral scales as $\lambda^\mu$ as $\lambda\to0$, one says the
degree of infrared divergence is~$\mu$.
For example, in $d$~dimensions the momentum-space volume element~$d^dk$ 
has $\mu(d^dk)=d$.
If~$\mu>0$, an integral is infrared convergent.

The conclusions derived above for arbitrary diagrams apply equally
well to the one-particle irreducible ones contributing to the self
energy.
It is convenient to route the external momentum~$p$ along the ``main
line,'' the massive quark line that runs all the way through a
self-energy diagram.
Off the main line the momenta are independent of~$p$, and the
degrees of infrared divergence are straightforward.
Soft gluon and ghost propagators contribute $\mu(\Delta)=-2$, and
soft massless quark propagators $\mu(S)=-1$.
Soft three-gluon and gluon-ghost vertices contribute $\mu(V_3)=+1$, 
and other soft vertices $\mu(V)=0$.
In a closed loop the massive quark
propagator~$S_0(k)=1/(i\kern+0.1em /\kern-0.55em k +m_0)$
has degree of infrared divergence~$0$.

The internal parts of the main line have propagator~$S_0(p+k)$.
When~$k$ is soft
\begin{equation}
S_0(p+k)=
\frac{1}{i(\kern+0.1em /\kern-0.55em p+\kern+0.1em /\kern-0.55em k)+m_0}
\to\frac{m_0-i\kern+0.1em /\kern-0.55em p}{p^2+2p\cdot k+m_0^2}.
\end{equation}
Off shell (away from the branch point) such lines have degree of 
infrared divergence~$0$.
On shell, however,%
\footnote{With $p^2=-M^2$ one treats $m_0^2-M^2$ as higher order 
in~$g_0^2$.}
\begin{equation}
    S_0(p+k)\to
    \frac{m_0-i\kern+0.1em /\kern-0.55em p}{2p\cdot k},
\end{equation}
which gives degree~$-1$.

When all loop momenta are soft and $p^2=-M^2$, an arbitrary QCD (or 
QED!) self-energy diagram~$G_L$ with $L$ loops, but no massive quark 
loops, has degree of infrared divergence
\begin{equation}
    \mu(G_L)=1+L(d-4)
    \label{degree}
\end{equation}
in $d$ dimensions.
This holds at one loop.
Higher-loop diagrams can be built by adding gluons, ghost loops, or
(for now) massless quark loops.
It is enough to insert the loops into gluon propagators; more gluons can
be added later.
Loop insertions give $\mu(G_{L+1})=\mu(G_L)+d-4$.
New gluons give $\mu(G_{L+1})=\mu(G_L)+d-2+\mu_1+\mu_2$, where $d$
comes from the new loop, $-2$ from the gluon propagator, and the
$\mu_i$ are degrees associated with each end.
The ends can be on a gluon or ghost line: $\mu_i=1-2=-1$ for vertex
and propagator; on a three-gluon vertex: $\mu_i=-1$ from changing
the three- to a four-gluon vertex; on the main line or a massless
quark: $\mu_i=-1$ for the propagator.%
\footnote{When the massive quark is off shell, the degree is
increased by the number of main-line propagator segments---even
more convergent.}
Thus, these ways all give $\mu(G_{L+1})=\mu(G_L)+d-4$.
Replacing an internal $n$-vertex polygon of a massless quark with a
massive one increases~$\mu(G_L)$ by~$n$.
Therefore, in four dimensions no infrared divergences can arise
from the region with all loop momenta soft.

Infrared divergences may come, however, from regions with some loop
momenta soft and others not.
According to the physical picture of the self energy, the non-soft
lines can be shrunk to a point, augmenting the foregoing analysis
with composite vertices.
For $n>3$, $n$-point vertices~$V_n$ contribute $\mu(V_n)=0$ to the
total.
Additional soft gluons attached to such vertices come with a 
propagator and a loop integration, adding $d-2$ to the total degree.
Composite three-point vertices have the same infrared power counting as
their fundamental counterparts.
For example, gauge invariance guarantees the beneficial $\mu(V_3)=+1$
for the (composite) vertex of three soft gluons, in the same way it
safeguards renormalizability.
Thus, multi-point composite vertices do not lower the degree of 
infrared divergence in four dimensions.

Internal, hard self-energy diagrams shrink to two-point vertices.
In massless quark loops, the two-point vertex and extra
propagator yield the harmless factor
$\Sigma(k)/(i\kern+0.1em /\kern-0.55em k)=A(k)$, as required by
chiral symmetry.
In massive quark loops the additional factor $\Sigma(k)/m_0\to B(k)$ is
also harmless.
(These self energies are off shell and, therefore, well behaved.)
Gauge symmetry provides two powers of soft momenta at two-point gluon
and two-point ghost vertices, cancelling the extra propagator.
Thus, these two-point vertices do not pose a problem.

\begin{figure}[btp]
    \epsfxsize=\textwidth 
    \epsfbox{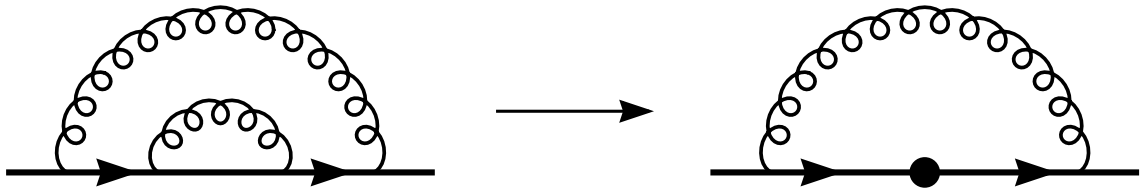}
    \caption{Origin of the infrared divergence at two loops.}
    \label{fig:2loop}
\end{figure}

What remains are two-point vertices of the massive quark on the main
line.
The extra factor from inserting such a two-point vertex is
$\Sigma(p,m_0)[m_0-i\kern+0.1em /\kern-0.55em p]/2p\cdot k$, which
lowers the degree of infrared divergences to~$\mu\le 0$.
For example, at two loops it is known~\cite{Tar81} that the self energy
$\Sigma^{[2]}(-m_0^2,m_0)$ is infrared divergent.
The origin of the divergence, as the above analysis implies, is sketched
in Fig.~\ref{fig:2loop}.
The problem worsens at higher and higher orders, as more and more
two-point vertices can arise on the main line.

\subsection{Infrared Cancellation}
For the pole mass at two loops the infrared divergence in
$\Sigma^{[2]}(-m_0^2,m_0)$ is cancelled by the $O(g_0^4)$ part of
$\Sigma^{[1]}(-M^2,m_0)$.
To examine this mechanism in general it is convenient to solve
for the bare mass that implies a desired pole mass, namely
\begin{equation}
    m_0=MZ_m^{-1},   \label{m_0}
\end{equation}
with
\begin{equation}
    Z_m^{-1}=\frac{1-A(-M^2,M/Z_m)}{1-B(-M^2,M/Z_m)}
    \label{Z-1}
\end{equation}
showing the iterative nature of the solution.
The $L$th coefficient~$[Z_m^{-1}]^{[L]}$ of the iterated expansion
is infrared finite if and only if the coefficients~$Z_m^{[l]}$,
$l\le L$, are infrared finite.

The coefficients of the iterated expansion involve certain combinations
of the coefficients of the self energy functions.
Let
\begin{equation}
    \Sigma(p^2,MZ_m^{-1})=\sum_{l}g_0^{2l}\bar{\Sigma}^{[l]}(p^2,M)
\end{equation}
define the coefficients~$\bar{\Sigma}^{[l]}(p^2,M)$ [and, by 
implication, $\bar{A}^{[l]}(p^2,M)$ and~$\bar{B}^{[l]}(p^2,M)$].
Iterative expansion yields
\begin{equation}
    \bar{\Sigma}^{[l]}(p^2,M)=\Sigma^{[l]}(p^2,M) +
    \sum_{i=1}^{l-1}\sum_{j=1}^{i}
    \frac{1}{j!}\left[(Z_m^{-1}-1)^j\right]^{[i]}M^j \left.
    \frac{\partial^j\Sigma^{[l-i]}}{\partial m_0^j}\right|_{m_0=M}.
  \label{bar Sigma}
\end{equation}
% \begin{equation}
%   \begin{array}{l}
%     \bar{\Sigma}^{[l]}(p^2,M)=\Sigma^{[l]}(p^2,M) + {}\\
%     \hspace{2.0em}\displaystyle
%     \sum_{i=1}^{l-1}\sum_{j=1}^{i}
%     \frac{1}{j!}\left[(Z_m^{-1}-1)^j\right]^{[i]}M^j \left.
%     \frac{\partial^j\Sigma^{[l-i]}}{\partial m_0^j}\right|_{m_0=M}.
%   \end{array}
%   \label{bar Sigma}
% \end{equation}
Explicit calculation shows that 
$\bar{\Sigma}^{[1]}(p^2,M)=\Sigma^{[1]}(p^2,M)$ is infrared finite 
and, when $p^2=-M^2$, gauge independent.
At two loops $\bar{\Sigma}^{[2]}(-M^2,M)$ is also infrared finite and 
gauge independent, even though $\Sigma^{[2]}(-M^2,M)$ is 
not~\cite{Tar81,Gra90}.

Since
\begin{eqnarray} {}
    [Z_m^{-1}]^{[1]}&=&\bar{B}^{[1]} - \bar{A}^{[1]}, \\
{}  [Z_m^{-1}]^{[2]}&=&\bar{B}^{[2]} - \bar{A}^{[2]} +
    [Z_m^{-1}]^{[1]} \bar{B}^{[1]},
\end{eqnarray}
with all self-energy functions evaluated at $p^2=-M^2$ and $m_0=M$,
one has a basis for a proof by induction.
Let us assume that the $[Z_m^{-1}]^{[l]}$, $l<L$, are infrared finite.
To show that $[Z_m^{-1}]^{[L]}$ is also infrared finite it is enough
to show that $\bar{\Sigma}^{[L]}(-M^2,M)$ is infrared finite.
The induction hypothesis implies that $\bar{\Sigma}^{[l]}(-M^2,M)$, 
$l<L$, is infrared finite (otherwise $[Z_m^{-1}]^{[l]}$ would not be).

\begin{figure}[btp]
    \epsfxsize=\textwidth 
    \epsfbox{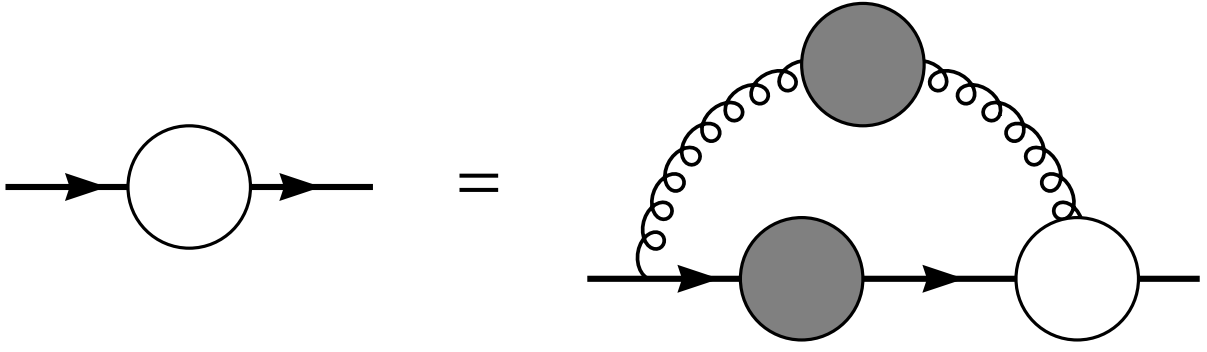}
\caption[fig:DSeq]{Dyson-Schwinger equation for the self energy.
Grey blobs denote full propagators and white blobs denote
one-particle irreducible functions.}
\label{fig:DSeq}
\end{figure}

It would be a nightmare to identify all infrared divergences and
verify cancellation on the right-hand side of Eq.~(\ref{bar Sigma}).
Instead, it is more efficient to study $\bar{\Sigma}^{[L]}(p^2,M)$
directly.
The Dyson-Schwinger equation for the self-energy, depicted in
Fig.~\ref{fig:DSeq}, is a useful tool.
Power-counting and the induction hypothesis together say that the only 
new infrared divergence at~$L$ loops can come from diagrams with the 
$(L-1)$th-order expansion of the quark propagator and an additional 
gluon:
\begin{equation}
    \bar{\Sigma}^{[L]}_{\text{IR}}=\int\frac{d^dk}{(2\pi)^4}
    \frac{\delta_{\mu\nu}-\xi k_{\mu}k_{\nu}/k^2}{k^2}
    \gamma_\mu[S(p+k)]^{[L-1]}\gamma_\nu.
    \label{DSIR}
\end{equation}
Diagrams with a higher-order 1PI vertex function or gluon propagator 
would have a quark propagator with $L-2$ (or fewer) loops; they can be 
infrared divergent only if $\bar{\Sigma}^{[l]}$, for some $l<L$, were 
too---contrary to the induction hypothesis.

The key to obtaining $\bar{\Sigma}^{[L]}_{\text{IR}}$ from the 
right-hand side of Eq.~(\ref{DSIR}) is to write ($q=p+k$)
\begin{equation}
    S(q)=\frac{1}{i\kern+0.1em /\kern-0.55em q + M -
    \left[\Sigma(q,MZ_m^{-1})-M(Z_m^{-1}-1)\right]},
    \label{S(q)}
\end{equation}
treating $i\kern+0.1em /\kern-0.55em q + M$ as $O(g_0^0)$ and 
expanding the bracket (iteratively) in~$g_0^2$.
This expansion, for $p^2=-M^2$ and~$k$ soft, produces a sum of chains
\widetext
\begin{equation}
    [S(p+k)]^{[L-1]}= \frac{M-i\kern+0.1em /\kern-0.55em p}{2p\cdot k}
    \sum\prod_j \cdots
    \left[\bar{\Sigma}(p+k,M)-M(Z_m^{-1}-1)\right]^{[l_j]}
    \frac{M-i\kern+0.1em /\kern-0.55em p}{2p\cdot k}\cdots\cdot
    \label{propagator}
\end{equation}
In any term of the sum the factors' superscripts~$l_j$ add up 
to~$L-1$, and the sum is over all such partitions of~$L-1$.
Since~$k$ is soft the quantity in brackets reduces to
\begin{eqnarray}
    \left[\bar{\Sigma}(p+k,M)-M(Z_m^{-1}-1)\right]^{[l]} & \to &
    \left[i\kern+0.1em /\kern-0.55em p A(p^2,M/Z_m) +
    MZ_m^{-1}B(p^2,M/Z_m)-M(Z_m^{-1}-1)\right]^{[l]} \nonumber \\
    & = &
    (i\kern+0.1em /\kern-0.55em p +M)\bar{A}^{[l]}(-M^2,M) + O(k).
    \label{bracket}
\end{eqnarray}
\narrowtext\noindent
The second step follows from Eq.~(\ref{Z-1}) and setting~$p^2\to-M^2$.
In Eq.~(\ref{propagator}) the right-hand side of Eq.~(\ref{bracket}) 
multiplies $M-i\kern+0.1em /\kern-0.55em p$; the $O(1)$ part of the 
product vanishes for $p^2=-M^2$, leaving a remainder of $O(k)$.%
\footnote{One might worry whether the remainder is an 
infrared-divergent derivative of the self energy.
But because~$p+k$ is off shell by~$k$, the remainder is proportional
to~$k\ln k$.}
Consequently, $[S(p+k)]^{[L-1]}$ has degree of infrared 
divergence~$-1$, just like $S_0(p+k)$.
Thus, $\bar{\Sigma}^{[L]}(-M^2,M)$ is infrared-finite, as was to be 
proved.

Although the above formulae are a bit clumsy, the mechanism that 
cancels the infrared divergences is simple.
Equation~(\ref{S(q)}) says to split the bare mass into the pole mass
plus a counterterm.
The counterterm, like the shrunken self-energy subdiagram, produces a 
two-point vertex.
Infrared divergences cancel in the (next order's) pole mass, because 
the combination of the two does not degrade the infrared power 
counting.
In QED, this mechanism was identified in a footnote to
Ref.~\cite{Nak58}.

\section{Gauge Independence}
\label{GI}
The gauge invariance of the mass renormalization factor~$Z_m$ is 
``nearly obvious.''
If the ultraviolet regulator respects gauge symmetry, the bare mass 
has a gauge-invariant meaning.
  From a physical point of view it would be unsettling if the pole mass 
were to depend on the gauge.
Thus, the ratio $Z_m=M/m_0$ ought to be gauge invariant too.
Without the infrared-finiteness established above, however, a proof 
would require painstaking separation of infrared and ultraviolet
regulators.

With infrared-finiteness one can study the gauge dependence by treating
the massive quark like a normal particle, as long as one discusses
perturbation theory only.
For~$p^2$ near~$-M^2$ the propagator takes the form
\begin{equation}
    S(p)=\frac{Z_2}{i\kern+0.1em /\kern-0.55em p +M},
    \label{near pole}
\end{equation}
where $Z_2$ is the field renormalization factor.%
\footnote{$Z_2$ is not infrared divergent when~$p^2$ is close to, but 
not equal to,~$-M^2$.}
Because the pole mass is only defined perturbatively, $M$ denotes 
here the pole mass through some finite order in perturbation theory.

The previous section assumed a gauge fixing term 
$(\lambda/2)\int(\partial\cdot A)^2d^4x$.
[In Eq.~(\ref{DSIR}), $\xi=1-\lambda^{-1}$.]
Suppose~$M$ depends on~$\lambda$.
A shift~$\Delta\lambda$ induces a first-order change
\begin{equation}
    \Delta S(p)=\frac{\Delta\lambda}{i\kern+0.1em /\kern-0.55em p +M}
    \left[\frac{\partial Z_2}{\partial\lambda} -
    \frac{\partial M}{\partial\lambda}
    \frac{Z_2}{i\kern+0.1em /\kern-0.55em p +M} \right].
    \label{double pole}
\end{equation}
Note that $\partial M/\partial\lambda$ multiplies a double pole.

On the other hand, the propagator is given by
\begin{equation}
    S(p)=\text{FT}
    \langle 0|T\psi(x)\bar{\psi}(y)|0 \rangle.
    \label{two-point}
\end{equation}
  From Eq.~(\ref{two-point}), the change is 
\begin{equation}
    \Delta S(p)=\frac{\Delta\lambda}{2}\text{FT}\int d^dz
    \langle 0|T\psi(x)\bar{\psi}(y)(\partial\cdot A(z))^2|0 \rangle
    \label{change}
\end{equation}
A double pole would develop if the quark could scatter off (two)
gluons with scalar polarization, which cannot happen because the
scalar polarization decouples from the physical state space.
The mass~$M$ cannot, therefore, depend on the gauge parameter, 
although the residue~$Z_2$ certainly can.

(Conversely, one can see immediately that the insertions generated by 
a shift in the bare mass or gauge coupling {\em would\/} develop 
a double pole and, thus, a shift in the pole mass.)

For a more general gauge-fixing function~$f^a(A(z))$
(and change~$\Delta f^a$) the argument is similar.
Let~$s$ denote the BRS operator and~$\eta$ ($\bar{\eta}$) the 
(anti)-ghost field.
The change in the two-point function involves~\cite{Wei93}
\widetext
\begin{eqnarray}
    \langle s[\bar{\eta}_a(z)\Delta f^a(z)] \psi(x)\bar{\psi}(y)\rangle
    & \sim & 
    \langle \bar{\eta}_a(z)\Delta f^a(z) s[\psi(x)\bar{\psi}(y)]\rangle
    \\ & \sim &
    \langle\bar{\eta}_a(z)\Delta f^a(z)t^b[\eta_b(x)-\eta_b(y)]
    \psi(x)\bar{\psi}(y)\rangle
    \label{ghost}
\end{eqnarray}
\narrowtext\noindent
As before, but now because the ghosts decouple, this expression cannot 
develop a double pole.

\section{Conclusions}
\label{Conclusions}

The pole mass is widely used in the phenomenology of QCD and, when
quark momenta are small compared to the mass, in nonrelativistic
QCD and heavy-quark effective theory.
In many of these contexts it is natural: it has considerable
intuitive appeal, and it can be calculated with any ultraviolet
regulator and in any effective theory.
In some circles the pole mass---as an experimental quantity---has
rightly fallen into disfavor, because infrared renormalons obstruct
an unambiguous determination~\cite{Big94,Ben94}.
Had the pole mass turned out to be either infrared divergent or gauge
dependent, one ought to have abandoned the pole mass for more basic
reasons.

Fortunately, Sec.~\ref{IRF} shows that the perturbative pole mass
is infrared finite, even though the on-shell self energy is not.
At higher orders the pole mass requires an iterative expansion and,
thus, (infrared divergent) derivatives of the self energy.
In the pole mass the total infrared divergence vanishes.
The cancellation mechanism and, indeed, the power counting in QCD
are the same as in standard QED.
Attaching a virtual photon to an on-shell massive electron line has
the same effect as attaching a gluon anywhere in a massive-quark
self-energy diagram.
I cannot imagine that infrared-finiteness of the electron mass has
never been proved, but, except for a footnote~\cite{Nak58},
I have not found a reference with a proof.

Because of its physical appeal, the pole mass remains valuable
theoretically.
In addition to matching to effective theories, mentioned above, it is
similarly useful in relating lattice QCD to continuum renormalization
schemes~\cite{Mer98}.
A example of considerable phenomenological interest is the
application of (NR)QCD to threshold production of heavy quarks.
There the pole mass is nearly irresistible, but it has been pointed out
recently that infrared sensitivity in the mass's definition is conferred
on the QCD potential as well~\cite{Hoa98,Ben98,Ura98}.
Indeed, Ref.~\cite{Ben98} notes that a formula equivalent to
Eq.~(\ref{bar Sigma}) would be helpful in showing that infrared
divergences in the pole mass cancel at every order.

\acknowledgments
An anonymous referee of Ref.~\cite{Mer98} pointed out that the
infrared-finiteness and gauge-independence of the pole mass had not been
proved in QCD beyond two loops.
I~thank Tony Duncan and George Sterman for helpful conversations
and Aida El-Khadra, Aneesh Manohar, and Scott Willenbrock for
comments on the manuscript.
I~would also be grateful to anyone who would inform me of a prior
(published) proof, even in QED.
Fermilab is operated by Universities Research Association Inc.,
under contract with the U.S. Department of Energy.\\[1.0em]

\noindent
{\em Note added in proof:\/}
The gauge independence (but not the infrared finiteness) of the pole
mass has been proven in QED~\cite{Atk79,Bre95} and
QCD~\cite{Bre95,Joh86};
I thank V. Miransky, T. Steele, and G. Kilcup, respectively, for drawing
my attention to these works.
A proof of both infrared finiteness and gauge independence of the
electron mass to all orders in QED is given in Ref.~\cite{Bro92};
I thank A. Schreiber for drawing my attention to this text.

\end{document}